\documentclass[11pt,a4paper,numbers,sort&compress]{article}
\usepackage[utf8]{inputenc}
\pdfoutput=1
\usepackage{jheppub}
\usepackage{graphicx} 
\usepackage{amsmath,amsfonts,amssymb}
\usepackage{physics}

\newcommand{\ul}[1]{\underline{#1}}
\newcommand{\ol}[1]{\overline{#1}}
\newcommand\nn{\nonumber}

\preprint{LCTP-24-09}

\title{The fate of boundary terms in dimensional reductions}

\author{Robert J. Saskowski}
\emailAdd{rsaskows@umich.edu}

\affiliation{Leinweber Center for Theoretical Physics, Randall Laboratory of Physics\\The University of Michigan, Ann Arbor, MI 48109-1040, USA}
\affiliation{Center for Joint Quantum Studies and Department of Physics,\\
School of Science, Tianjin University, Tianjin 300350, China}

\abstract{Gibbons-Hawking-York (GHY) terms are typically neglected when performing dimensional reductions of gravitational theories. We consider the reduction of such terms for both two-derivative and four-derivative theories in general dimensions. We demonstrate a robust consistency wherein the GHY term in the original, higher-dimensional theory translates directly to the appropriate GHY term in the dimensionally reduced theory. In particular, this gives a novel way of generating such terms for higher-derivative corrections. We carry out this procedure for Gauss-Bonnet, Chern-Simons modified, and $f(R)$ gravities to derive novel boundary terms.}

\date{\today}

\begin{document}
\maketitle

\section{Introduction and summary}
The idea of dimensionally reducing a gravitational theory was considered nearly a century ago by Kaluza~\cite{Kaluza:1921tu} and Klein~\cite{Klein:1926tv}. Such reductions have since found broad application from string theory compactifications~\cite{Candelas:1985en} to practical experiments~\cite{Obied:2023clp}. The classic example is the Scherk-Schwarz reduction \cite{Scherk:1979zr}, wherein the internal space is taken to be a group manifold which becomes the gauge group of the effective lower-dimensional theory. In general, one gets an infinite tower of massive states controlled by the size of the compact internal dimensions. Much work has been done on classifying for which reductions it is consistent to truncate this tower, both in terms of the allowed geometries (see \emph{e.g.}~\cite{Duff:1984hn,Cvetic:2000dm,Cvetic:2003jy,Lee:2014mla}) and the effect of higher-derivatives (see \emph{e.g.}~\cite{Chang:2021tsj,Liu:2023fqq,Liu:2023fmv}). However, in such studies, the emphasis is often placed exclusively on the bulk gravitational action, and integration by parts is performed without concern for the resulting boundary terms. This is acceptable if we are only interested in local data: equations of motion, conserved currents, symmetries, and Noether identities depend only on the Lagrangian density. Nevertheless, ignoring the boundary terms is equivalent to not specifying boundary conditions and makes the variational principle ill-posed.

It was shown by York~\cite{York:1972sj} and Gibbons and Hawking~\cite{Gibbons:1976ue} that in order for the Dirichlet variational problem to be well-defined, the bulk Einstein-Hilbert action,
\begin{equation}
    S_\mathrm{EH}=\frac{1}{2\kappa_D}\int_{\mathcal M}\dd[D]x \sqrt{-g}\,R\,,\label{eq:EHsec1}
    \end{equation}
must be supplemented by a Gibbons-Hawking-York (GHY) boundary term
\begin{equation}
    S_\mathrm{GHY}=\frac{1}{\kappa_D}\int_{\partial\mathcal M}\dd[D-1]x \sqrt{-h}\,K,\label{eq:GHYog}
\end{equation}
where $R$ is the Ricci scalar for the metric $g_{\mu\nu}$ on the spacetime manifold $\mathcal M$, $\kappa_D=8\pi G_N^{(D)}$ is the standard gravitational coupling in $D$ dimensions, $h_{ab}$ is the induced metric on the boundary $\partial\mathcal M$, and $K$ is the trace of the second fundamental form of $\partial\mathcal M$. This is because varying an action with second-order derivatives, as in \eqref{eq:EHsec1}, leads to terms schematically of the form $\partial\partial\delta g_{\mu\nu}$, which, after integration by parts, lead to boundary terms proportional to $\partial\delta g_{\mu\nu}\vert_{\partial\mathcal M}$. Such terms are not automatically removed by the Dirichlet boundary conditions,
\begin{equation}
    \delta g_{\mu\nu}\big\vert_{\partial\mathcal M}=0\,,
\end{equation}
so we must add the GHY term \eqref{eq:GHYog} to cancel them. This has since been generalized to a variety of higher-derivative theories of gravity, see \emph{e.g.}~\cite{Barth:1984jb,Teitelboim:1987zz,Myers:1987yn,Madsen:1989rz,Dyer:2008hb,Cremonini:2009ih,Guarnizo:2010xr,Smolic:2013gz,Bueno:2016dol,Teimouri:2016ulk,Deruelle:2017xel,Bueno:2018xqc,Lindstrom:2022mbe,Erdmenger:2022nhz}. There are also corresponding generalizations to more general boundaries (spacelike, null, non-smooth, \emph{etc}.)~\cite{Hartle:1981cf,Hayward:1993my,Parattu:2015gga,Parattu:2016trq,Chakraborty:2016yna,Cano:2018ckq,Brizuela:2023vmb} and to more general boundary conditions (Neumann, Robin, mixed, \emph{etc}.)~\cite{York:1986lje,Krishnan:2016mcj,Krishnan:2016tqj,Krishnan:2017bte}.

Such GHY terms are desirable as they ensure the path integral has the correct composition properties~\cite{Gibbons:1976ue} and the action is functionally differentiable~\cite{Padmanabhan:2014lwa}. They are also necessary to define conserved charges at infinity using the ADM prescription~\cite{Hawking:1995fd}, and these charges crucially depend on the choice of boundary conditions~\cite{Brown:1992br,Odak:2021axr}. For holographic theories, these ADM charges are identified with the charges of the dual field theory, and the GHY terms are necessary (but generally not sufficient) to regulate the IR divergences of the bulk action in holographic renormalization~\cite{Bianchi:2001kw}. Furthermore, the entire contribution to semiclassical black hole entropy comes from the GHY term~\cite{Brown:1992bq}. As such, it is important to understand the role of boundary terms.

The dimensional reduction of an action with a well-posed variational principle should lead to a reduced action with a well-posed variational principle. Hence, it is natural to expect that the dimensional reduction of a GHY term should yield the corresponding GHY term for the lower-dimensional theory. Indeed, it was observed in~\cite{Mueller-Hoissen:1989bdj} that the circle reduction of the GHY term \eqref{eq:GHYog} for the Einstein-Hilbert action yields precisely the GHY term for the lower-dimensional theory. Refs.~\cite{Svesko:2022txo,Gukov:2022oed} found similar results in sphere reductions to 2d, although they kept only one KK scalar and omitted the KK gauge field. Likewise, Refs.~\cite{Geng:2022slq,Geng:2022tfc} demonstrated this for a particular braneworld reduction from 3d to 2d in the context of wedge holography. See also~\cite{Edery:2009vr} for an interesting example of how boundary conditions can play an important role in dimensional reduction in a non-gravitational context. One motivation for us is to show that this continues to hold for more general reductions of gravitational theories. A second motivation is that we are interested in \emph{constructing} GHY terms. The procedure of computing GHY terms for theories is oftentimes tedious, especially for higher-derivative theories, since one must vary the action and determine what boundary terms will give the same answer after variation. We propose that dimensional reduction is often a straightforward approach to obtaining such terms.

As we will see, the story for two-derivative theories is rather uneventful; the extra terms that arise on the boundary from dimensional reduction are always canceled by equal and opposite terms that arise from integrating the bulk action by parts. In particular, we perform a group manifold reduction of the Einstein-Hilbert action and its GHY term on a unimodular Lie group and show that the reduction simply amounts to
\begin{equation}
    \frac{1}{\kappa_D}\int_{\partial\mathcal M}\dd[D]x\sqrt{-\hat h}\,\hat K\to \frac{1}{\kappa_d}\int_{\partial\mathcal B}\dd[d]x\sqrt{- h}\,K\,.
\end{equation}
This is simply the statement that the only term in the two-derivative action that needs a GHY term is the Einstein-Hilbert term, so no interesting new terms arise. 

However, at the four-derivative level, we use this procedure to generate new GHY terms.  In particular, dimensionally reducing the Gauss-Bonnet action allows us to easily show that the combination
\begin{equation}
    \int_{\mathcal M}\dd[D]x\sqrt{-g}\,R_{\mu\nu\rho\sigma}U^{\mu\nu\rho\sigma}-4\int_{\partial\mathcal M}\dd[D-1]x\sqrt{-h}\,U^{\mu\nu\rho\sigma}n_\mu K_{\nu\rho}n_\sigma
\end{equation}
leads to a well-posed Dirichlet variation of the metric, where $U_{\mu\nu\rho\sigma}$ is any tensor with the symmetries of the Riemann tensor that does not contain derivatives of the metric. We also demonstrate that the procedure can be used in reverse. Knowing only the answer for $D=4$ \cite{Grumiller:2008ie}, we show that the term
\begin{equation}
    \int_{\mathcal M}\Tr[R\land R]\land A_{(D-4)}+4\int_{\partial\mathcal M}\mathrm{CS}(K)\land A_{(D-4)}
\end{equation}
has a well-posed Dirichlet variation, where $A_{D-4}$ is a $(D-4)$-form field and $\mathrm{CS}(K)$ is the abelian Chern-Simons term for the extrinsic curvature. Finally, we use the dimensional reduction of $f(R)$ gravity to derive the appropriate GHY term for $f(R,T,\mathcal L_\mathrm{matter})$ gravity:
\begin{equation}
    \int_{\mathcal M}\dd[D]x\sqrt{-g}\,f(R,T,\mathcal L_\mathrm{matter})+2\int_{\partial\mathcal M}\dd[D-1]x\sqrt{-h}\,\partial_Rf(R,T,\mathcal L_\mathrm{matter})K\,,
\end{equation}
where $T$ is the trace of the stress-energy tensor. Of these three main results, the first was shown by~\cite{Barvinsky:1995dp}, while the other two are novel as far as we know.

The rest of this paper is organized as follows. In Section \ref{sec:twoDer}, we show the two-derivative group manifold reduction gives precisely the lower-dimensional GHY term. In Section \ref{sec:fourDeriv}, we investigate the effects of adding higher-derivative corrections and their corresponding GHY terms. In particular, we consider the Gauss-Bonnet term, the $A\land \Tr R\land R$ term, and $f(R)$ gravity. We conclude in Section \ref{sec:disc} and discuss future directions. Some technical details are relegated to the appendices.

\subsection*{Conventions}
We use a mostly positive metric signature $(-,+,+,\cdots,+)$. 
We take the Riemann and Ricci tensors to be given by 
$$R^{\mu}{}_{\nu\rho\sigma}=\partial_\rho\Gamma^\mu_{\sigma\nu}-\partial_\sigma\Gamma^\mu_{\rho\nu}+\Gamma^\mu_{\rho\lambda}\Gamma^\lambda_{\sigma\nu}-\Gamma^\mu_{\sigma\lambda}\Gamma^\lambda_{\rho\nu}\,,$$
and $R_{\mu\nu}=R^\rho{}_{\mu\rho\nu}$, respectively. We choose to weight our antisymmetrizations as $A_{[\mu\nu]}=\tfrac{1}{2}(A_{\mu\nu}-A_{\nu\mu})$ and our symmetrizations as $A_{(\mu\nu)}=\tfrac{1}{2}(A_{\mu\nu}+A_{\nu\mu})$. $\epsilon_{\mu\nu\cdots}$ is always taken to represent the Levi-Civita tensor and never the tensor density. The boundary is always assumed to be a smooth timelike hypersurface, and we take the normal to be outward-pointing and have unit norm.

\section{Two-derivative reduction}\label{sec:twoDer}
In this section, we consider a generic two-derivative theory of gravity with action
\begin{equation}
    S=\frac{1}{2\kappa_D}\int_\mathcal{M}\dd[D]x\sqrt{-\hat g}\,\qty(\hat R+\hat{\mathcal L}{}_\mathrm{matter})+\frac{1}{\kappa_D}\int_{\partial\mathcal M}\dd[D-1]x\sqrt{-\hat h}\,\hat K,
\end{equation}
where $\hat{\mathcal L}_\mathrm{matter}$ is some unspecified matter Lagrangian, which may contain a cosmological constant. We will follow the convention that hats denote $D$-dimensional fields. We will use early capital Latin letters for $D$-dimensional rigid indices and late capital Latin letters for $D$-dimensional curved indices. Our index-splitting convention is summarized as
\begin{equation*}
    M\to \{\mu,i\}\,,\qquad A\to \{\alpha, a\}\,.
\end{equation*}
We will treat the reduction of the bulk Einstein-Hilbert action and the boundary GHY term as two steps.

\subsection{Einstein-Hilbert action}
Let us do a standard group manifold reduction on a Lie group $\mathcal G$, with metric\footnote{Note that we are working in the ``string'' frame. This is done for computational convenience and will not affect the end conclusion, as we may always perform a Weyl rescaling of the metric.}~\cite{Scherk:1979zr}
\begin{equation}
    \dd s^2=g_{\mu\nu}(x)\dd x^\mu\dd x^\nu+g_{ij}(x)\eta^i\eta^j,\qquad \eta^i=\sigma^i+A^i,\label{eq:gpMfldRed}
\end{equation}
such that the spacetime manifold $\mathcal M$ becomes a fibration, $\mathcal G\to\mathcal M\to \mathcal B,$ over a base manifold $\mathcal B$. Here the coordinates $\hat x^M$ correspond to $\mathcal M$, $x^\mu$ to $\mathcal B$, and $y^i$ to $\mathcal G$. $g_{\mu\nu}$ is naturally the metric on $\mathcal B$ and $g_{ij}$ is a matrix of scalars charged under $\mathcal G$. $A^i$ is a principal $\mathcal G$-connection, with curvature given locally by
\begin{equation}
    F^i=\dd A^i-\frac{1}{2}f{}^i{}_{jk}A^j\land A^k,
\end{equation}
where $f^i{}_{jk}$ are the structure constants of $\mathfrak g=\mathrm{Lie}(\mathcal G)$. The gauge field naturally has an associated gauge-covariant derivative, which we denote as $D$. Given a $\mathfrak g$-valued form $t^i$, $D$ acts as
\begin{equation}
        Dt^i=\dd t^i-f^{i}{}_{jk}A^j\land t^k.
\end{equation}
The $\sigma^i$ are the left-invariant one-forms of $\mathcal G$, satisfying the Maurer-Cartan equation,
\begin{equation}
    \dd \sigma^i=-\frac{1}{2}f^i{}_{jk}\sigma^j\land\sigma^k\,,
\end{equation}
and generate right isometries of the metric. We may define the Cartan-Killing metric as
\begin{equation}
    \kappa_{ij}=-\frac{1}{2}f^k{}_{\ell i}f^\ell{}_{kj}\,,
\end{equation}
which is used to lower $\mathfrak{g}$ indices. We will not assume that $\mathcal G$ is semisimple and hence $\kappa_{ij}$ may be degenerate (or even vanishing). Hence, the inverse $\kappa^{ij}$ is generally not well-defined. However, we will assume that $\mathcal G$ is unimodular, as this is required to get a consistent truncation~\cite{Scherk:1979zr}, and so
\begin{equation}
    f^{i}{}_{ij}=0\,.
\end{equation}

Let $e^\alpha$ denote the vielbein associated to $g_{\mu\nu}$ and $e^a$ the vielbein associated with $g_{ij}$. A natural choice of vielbein is
\begin{equation}
        \hat e{}^\alpha=e^\alpha,\qquad \hat e{}^a=e^a_i\eta^i\,,
\end{equation}
such that $\dd s^2=\eta_{\alpha\beta}\hat e^\alpha\hat e^\beta+\delta_{ab}\hat e^a\hat e^b$. The corresponding spin connection on $\mathcal M$ is obtained from the first Cartan structure equation
\begin{equation}
    \dd\hat e^A+\hat\omega^A{}_B\land\hat e^B=0\,,
\end{equation}
and has components
\begin{align}
        \hat\omega{}^{\alpha\beta}&=\omega^{\alpha\beta}-\frac{1}{2}e_i^a F_{\alpha\beta}^i \hat e{}^a,\nn\\
        \hat\omega{}^{\alpha b}&=-P_\alpha{}^{bc}\hat e{}^c-\frac{1}{2}e_i^b F^i_{\alpha\beta}\hat e{}^\beta,\nn\\
        \hat\omega{}^{ab}&=Q_\alpha{}^{ab} \hat e{}^\alpha+\frac{1}{2}\mathcal C_{c,ab}\hat e{}^c,\label{eq:spinConn}
\end{align}
where $\omega$ denotes the spin connection associated with $\mathcal B$, and we have defined
\begin{align}
        P_\alpha{}^{ab}&=e^{i(a} D_\alpha e^{b)}_i,\nn\\
        Q_\alpha{}^{ab}&=e^{i[a} D_\alpha e^{b]}_i,\nn\\
        \mathcal C_{c,ab}&=f^i{}_{jk}\qty[e^c_i e^j_a e^k_b+e^b_i e^j_a e^k_c-e_i^a e^j_b e^k_c].
\end{align}
$Q$ naturally behaves as a composite $\mathcal G$-connection. It is straightforward to compute the Riemann tensor from the second Cartan structure equation,
\begin{equation}
    \hat R^{AB}=\dd\hat\omega{}^{AB}+\hat\omega{}^A{}_C\land\hat\omega{}^{CB}.
\end{equation}
Correspondingly, the Ricci tensor components are
\begin{align}
    \hat R{}_{\alpha\beta}&=R_{\alpha\beta}-\frac{1}{2}F_{\alpha\gamma}^iF_{\beta\gamma}^j g_{ij}-D_\beta P_{\alpha cc}-P_{\alpha cd}P_{\beta}{}^{cd},\nn\\
    \hat R{}_{\alpha b}&=\frac{1}{2}D_{\gamma}\qty(e_i^b F_{\alpha\gamma}^i)+\frac{1}{2}e_i^aF_{\alpha\gamma}^ie^j_bD_\gamma e_j^a+\frac{1}{2}e_i^bF_{\alpha\gamma}^iP_{\gamma cc}\,,\nn\\
    \hat R{}_{ab}&=-D^\gamma P_{\gamma ab}-P_{\gamma ad}\qty(e^i_bD_\gamma e^d_i)+\frac{1}{4}e_i^ae_j^bF_{\gamma\delta}^iF_{\gamma\delta}^j+P_{\gamma db}Q_{\gamma da}-2P_{\gamma a[b|}P_{\gamma c|c]}\nonumber\\
        &\qquad-\frac{1}{4}\mathcal C_{f,ac}f^i{}_{jk}e_i^fe^{jb}e^{kc}.
\end{align}
The Ricci scalar is
\begin{equation}
    \hat R=R-\frac{1}{4}F^i_{\alpha\beta}F^j_{\alpha\beta}g_{ij}-2\nabla^\gamma P_{\gamma cc}-P_{\gamma cc}P^{\gamma dd}-(P_\gamma{}^{ab})^2-V\,,\label{eq:Rs}
\end{equation}
where
\begin{equation}
    V=\frac{1}{4}\qty(f^i{}_{jk}f^\ell{}_{mn}g_{i\ell}g^{jm}g^{kn}+2f^i{}_{jk}f^j{}_{i\ell}g^{k\ell})
\end{equation}
is the scalar potential. The metric determinant decomposes as
\begin{equation}
    \sqrt{-\hat g}=\sqrt{-g}\sqrt{\det g_{ij}}\,.\label{eq:sdetg}
\end{equation}

Now, we plug \eqref{eq:Rs} and \eqref{eq:sdetg} into the action and integrate by parts. Jacobi's formula implies that
\begin{equation}
    \partial_\gamma\sqrt{\det g_{ij}}=\sqrt{\det g_{ij}}\,P_{\gamma cc}\,.
\end{equation}
We are then left with the bulk gravitational action
\begin{align}
    S_\mathrm{EH}&=\frac{1}{2\kappa_d}\int_{\mathcal B}\dd[d]x\sqrt{-g}\sqrt{\det g_{ij}}\,\qty[R-\frac{1}{4}g_{ij}F^i_{\alpha\beta}F^j_{\alpha\beta}+(P_{\gamma cc})^2-(P_\mu{}^{ab})^2-V]\,,
\end{align}
at the cost of a boundary term
\begin{equation}
    -\frac{1}{\kappa_d}\int_{\partial\mathcal B}\dd[d-1]x\sqrt{-g}\sqrt{\det g_{ij}}\,n^\gamma P_{\gamma cc}\,,\label{eq:bdyTerm}
\end{equation}
where we have absorbed the factor arising from the volume of $\mathcal G$ into $\kappa_d$. Note that we will not be concerned with the reduction of $\hat{\mathcal L}_\mathrm{matter}$ since we are interested only in what happens to the GHY terms.

\subsection{Gibbons-Hawking-York term}\label{sec:GHY}
Now, we turn our attention to the boundary. Denote the spacelike outward-pointing unit normal as $\hat n_M$. As a matter of notation,\footnote{Although not the most standard choice of notation, this is necessary to keep clear the distinction between rigid vs curved, bulk vs boundary, and internal vs external indices.} we will use underbars to denote indices that have been projected onto the boundary using the first fundamental form,\footnote{Strictly speaking, $\hat h$ is only the induced metric after being pulled back to the boundary. As a standard abuse of language, we will generally not distinguish between bulk indices that have been projected and those that have been properly pulled back to the boundary.} $\hat h^A{}_B=\delta^A_B-\hat n^A\hat n_B$. That is,
\begin{equation}
    \mathfrak T_{\ul A}\equiv \hat h_A{}^B\mathfrak T_B\,,
\end{equation}
for any tensor $\mathfrak T$. Note that expressions of the form $\nabla_A \mathfrak T_{\ul B}$ should be interpreted as $\nabla_A(h_B{}^C \mathfrak T_{C})$ rather than $h_B{}^C\nabla_A\mathfrak T_{C}$. The second fundamental form of $\partial\mathcal M$ is given by the Lie derivative of $\hat h$ along $\hat n$:
\begin{equation}
    \hat K{}_{AB}=\frac{1}{2}\mathcal L{}_{\hat n}\hat h{}_{AB}\,.\label{eq:sff}
\end{equation}

The normal $\hat n_M$ should be independent of $y^i$, and, since $\mathcal G$ is closed, we will assume that
\begin{equation}
    \hat n^\alpha=n^\alpha\,,\qquad \hat n^a=0\,.
\end{equation}
This is automatically satisfied whenever $\partial\mathcal M$ is a level set, which is given by a constraint \mbox{$\mathfrak f(\hat x)=0$~\cite{Mueller-Hoissen:1989bdj}}. This is because the standard KK reduction takes $\mathfrak f$ to be independent of $y$, which leads to a boundary of the form $\partial\mathcal M=\partial\mathcal{B}\times\mathcal G$. So, the normal form is given by
\begin{equation}
    \hat n=\frac{\partial_\mu \mathfrak f}{\hat g^{\nu\rho}\partial_\nu\mathfrak f\,\partial_\rho\mathfrak f}\,\dd x^\mu\,,
\end{equation}
which has no components along the internal directions. Since $\hat g^{\mu\nu}=g^{\mu\nu}$, the normal remains unchanged under dimensional reduction. For example, boundaries defined by a radial cutoff $r=R_c$ are level sets.

Using \eqref{eq:spinConn} and \eqref{eq:sff}, we may compute the extrinsic curvature components
\begin{equation}
    \hat K_{\ul{\alpha\beta}}=K_{\ul{\alpha\beta}}\,,\qquad K_{\ul{\alpha b\vphantom{\beta}}}=\frac{1}{2}e_i^bF^i_{\gamma\ul{\alpha\vphantom{\beta}}}n^\gamma\,,\qquad\hat K_{\ul{ab\vphantom{\beta}}}=n^\gamma P_{\gamma\ul{ab\vphantom{\beta}}}\,,
\end{equation}
with trace
\begin{equation}
    \hat K=K+n^\gamma P_{\gamma cc}\,.
\end{equation}
Furthermore, since the normal has no component along $\mathcal G$, it is clear that the metric determinant decomposes as
\begin{equation}
    \sqrt{-\hat h}=\sqrt{-h}\sqrt{\det g_{ij}}\,.
\end{equation}
Thus, the GHY term becomes
\begin{equation}
    S_{\partial\mathrm{EH}}=\frac{1}{\kappa_d}\int_{\partial\mathcal B}\dd[d-1]x\sqrt{-h}\sqrt{\det g_{ij}}\,\qty(K+n^\gamma P_{\gamma cc})\,.
\end{equation}
Combining this with the boundary term \eqref{eq:bdyTerm}, we are left with just the $d$-dimensional GHY term:
\begin{equation}
    \frac{1}{\kappa_d}\int_{\partial\mathcal B}\dd[d-1]x\sqrt{-h}\sqrt{\det g_{ij}}\,K\,.
\end{equation}
Cancellations such as this were also noticed in~\cite{Mueller-Hoissen:1989bdj,Svesko:2022txo,Gukov:2022oed} for more specialized reductions.

At this point, it is interesting to remark that the dimensional reduction also implicitly reduces our boundary conditions, in the sense that the $D$-dimensional Dirichlet boundary condition for pure gravity,
\begin{equation}
    \delta \hat g_{MN}\big\vert_{\partial\mathcal M}=0\,,
\end{equation}
becomes the Dirichlet boundary conditions for the reduced theory,
\begin{equation}
    \delta g_{\mu\nu}\big\vert_{\partial\mathcal B}=0\,,\qquad \delta A^i_{\mu}\big\vert_{\partial\mathcal B}=0\,,\qquad \delta g_{ij}\big\vert_{\partial\mathcal B}=0\,.
\end{equation}
We also see that the number of degrees of freedom,
\begin{equation}
    \frac{d(d-3)}{2}+(d-2)(D-d)+\frac{(D-d)(D-d+1)}{2} =\frac{D(D-3)}{2}\,,
\end{equation}
remains unchanged, as we would expect.

\section{Four-derivative reduction}\label{sec:fourDeriv}
We now turn our attention to higher-derivative theories. As we saw in the previous section, we can dimensionally reduce GHY terms to obtain the GHY terms for lower-dimensional theories. We may put this to use as a procedure for \emph{generating} GHY terms. Of course, the addition of higher-derivative terms brings subtleties since the equations of motion are no longer two-derivative in general. The simplest approach is to specify additional boundary data, as will be the case for $f(R)$ gravity. This corresponds to the fact that higher-derivative theories generically have more degrees of freedom than their two-derivative cousins. Alternatively, we may work with actions that lead to two-derivative equations of motion, such as the Lovelock actions \cite{Lovelock:1970zsf,Lovelock:1971yv}. Other possibilities, which we will not explore here, include the use of auxiliary fields \cite{Cvetic:2001bk,Deruelle:2009zk,Altas:2024utf} as well as field redefinitions \cite{Cremonini:2009ih} to leverage the perturbative nature of higher-derivative corrections.

\subsection{Gauss-Bonnet}
We now consider the reduction of a Gauss-Bonnet term
\begin{equation}
    S_{\mathrm{GB}}=\frac{1}{2\kappa_D}\int_\mathcal{M}\dd[D]x\sqrt{-\hat g}\,\hat{\mathcal X}_4\,,\qquad \hat{\mathcal X}_4=\hat R^{ABCD}\hat R_{ABCD}-4\hat R^{AB}\hat R_{AB}+\hat R^2.\label{eq:GBinit}
\end{equation}
The appropriate GHY term is~\cite{Teitelboim:1987zz,Myers:1987yn}
\begin{align}
    S_\mathrm{\partial GB}&=\frac{1}{\kappa_D}\int_{\partial\mathcal M}\dd[D-1]x\sqrt{-\hat h}\,\hat{\mathcal Q}_3\,,\nn\\
    \hat{\mathcal Q}_3&=-\frac{2}{3}\hat K^3+2\hat K\hat K^{\ul{AB}}\hat K_{\ul{AB}}-\frac{4}{3}\hat K_{\ul{AB}}\hat K^{\ul{BC}}\hat K_{\ul{C}}{}^{\ul A}-4\qty(\hat{\mathcal R}_{\ul{AB}}-\frac{1}{2}\hat{\mathcal R}\hat h_{\ul{AB}})\hat K^{\ul{AB}}\,,\label{eq:GBGHYinit}
\end{align}
where $\hat{\mathcal R}_{\ul{AB}}$ and $\hat{\mathcal R}$ are the induced Ricci tensor and scalar on the boundary, respectively. Note that varying the action \eqref{eq:GBinit} (with the GHY term \eqref{eq:GBGHYinit}) results in equations of motion that are second-order in derivatives, and hence the metric does not propagate any additional degrees of freedom compared to Einstein gravity. As such, it is sufficient to specify only Dirichlet boundary conditions, $\delta \hat g_{MN}\vert_{\partial\mathcal M}=0$.

Since the GHY term is more involved than the preceding cases, and since we will be mainly interested in the curvature and field strength terms, we will use a simplified circle reduction ansatz\footnote{Of course, such an ansatz will never be a consistent truncation, but it suffices for constructing lower-dimensional GHY terms.}
\begin{equation}
    \dd s^2=g_{\mu\nu}\dd x^\mu\dd x^\nu+\qty(\dd z+A_\mu\dd x^\mu)^2.\label{eq:U1red}
\end{equation}
Hence, $A$ is now just a $U(1)$ gauge field and $d=D-1$. This will be sufficient to tell us the GHY terms, which we may then easily check. We again label the normal vector as $\hat n_M$ and assume $\hat n_\alpha=n_\alpha$, $\hat n_{\ol z}=0$. The vielbein is naturally
\begin{equation}
    \hat e^{\alpha}=e^\alpha\,,\qquad \hat e^{\ol z}=\dd z+A\,.
\end{equation}
The Riemann tensor is easily found to be
\begin{align}
    \hat R_{\alpha\beta\gamma\delta}&=R_{\alpha\beta\gamma\delta}-\frac{1}{2}F_{\alpha\beta}F_{\gamma\delta}-\frac{1}{2}F_{\alpha[\gamma|}F_{\beta|\delta]}\,,\nn\\
    \hat R_{\alpha\beta\gamma\ol z}&=-\frac{1}{2}\nabla_\gamma F_{\alpha\beta}\,,\nn\\
    \hat R_{\alpha\ol z\beta\ol z}&=\frac{1}{4}F_{\alpha\gamma}F_{\beta}{}^\gamma\,.\label{eq:Riemann}
\end{align}
The Ricci tensor components are thus
\begin{equation}
    \hat R_{\alpha\beta}=R_{\alpha\beta}-\frac{1}{2}F_{\alpha\gamma}F_{\beta\gamma}\,,\qquad\hat R_{\alpha\ol z}=\frac{1}{2}\nabla^\gamma F_{\alpha\gamma}\,,\qquad\hat R_{\ol{zz}}=\frac{1}{4}F^2\,,
\end{equation}
and the Ricci scalar is
\begin{align}
    \hat R=R-\frac{1}{4}F^2.
\end{align}
The Gauss-Bonnet action thus becomes
\begin{align}
    S_{\mathrm{GB}}&=\frac{1}{2\kappa_d}\int_\mathcal{B}\dd[d]x\sqrt{-g}\bigg( \mathcal X_4-\frac{3}{2}R_{\alpha\beta\gamma\delta}F^{\alpha\beta}F^{\gamma\delta}+4R_{\alpha\beta}F^{\alpha\gamma}F^\beta{}_{\gamma}-\frac{1}{2}RF^2+\qty(\nabla F)^2\nn\\
    &\kern9em-2\qty(\nabla^\alpha F_{\alpha\beta})^2+\frac{3}{16}(F^2)^2-\frac{3}{8}F^4\bigg)\,,
\end{align}
where $\mathcal{X}_4$ is the $d$-dimensional Gauss-Bonnet term. This may be integrated by parts to give
\begin{align}
    S_\mathrm{GB}&=\frac{1}{2\kappa_d}\int_\mathcal{B}\dd[d]x\sqrt{-g}\bigg( \mathcal X_4-\frac{1}{2}R_{\alpha\beta\gamma\delta}F^{\alpha\beta}F^{\gamma\delta}+2R_{\alpha\beta}F^{\alpha\gamma}F^\beta{}_{\gamma}-\frac{1}{2}RF^2+\frac{3}{16}(F^2)^2-\frac{3}{8}F^4\bigg)\nn\\
    &\quad+\frac{1}{\kappa_d}\int_{\partial\mathcal B}\dd[d-1]x\sqrt{-h}\qty(-n_\gamma F_{\alpha\beta}\nabla^\alpha F^{\beta\gamma}+n^\alpha F_{\alpha\beta}\nabla_\gamma F^{\beta\gamma})\,.\label{eq:GB}
\end{align}
Note that we have not used any equations of motion, so this expression is valid regardless of any other couplings we add to the action.

We may split this boundary term into its normal and tangential projections and rewrite the tangential bulk covariant derivatives $\nabla$ in terms of the boundary covariant derivative $\mathcal D$ (see Appendix \ref{app:math} for a summary of the relevant mathematics). The normal projections cancel, and we are left with
\begin{align}
    -n_\gamma F_{\alpha\beta}\nabla^\alpha F^{\beta\gamma}+n^\alpha F_{\alpha\beta}\nabla_\gamma F^{\beta\gamma}&=-n_\gamma F_{\alpha\beta}\mathcal D^\alpha F^{\beta\gamma}+n^\alpha F_{\alpha\beta}\mathcal D_\gamma F^{\beta\gamma}-K^{\ul{\alpha\beta}}F_{\ul{\alpha\gamma}}F_{\ul\beta}{}^{\ul\gamma}\nn\\
    &\quad+K^{\ul{\alpha\beta}}F_{\ul{\alpha\vphantom{\gamma}}\gamma}F_{\ul{\beta}\delta}n^\gamma n^\delta-KF_{\gamma\alpha}F_{\delta}{}^{\alpha}n^\gamma n^\delta\,.
\end{align}

Now we turn to the boundary. The metric \eqref{eq:U1red} leads to the second fundamental form components
\begin{equation}
    \hat K_{\ul{\alpha\beta}}=K_{\ul{\alpha\beta}}\,,\qquad K_{\ul{\alpha\vphantom{\beta}}\ol z}=\frac{1}{2}F_{\gamma\ul{\alpha\vphantom{\beta}}}n^\gamma\,,\qquad \hat K_{\ol{zz}}=0\,,\label{eq:exCurv}
\end{equation}
with trace
\begin{equation}
    \hat K=K\,.
\end{equation}

To find the induced Riemann tensor on the boundary, $\hat{\mathcal{R}}_{\ul{ABCD}}$, we may use the Gauss equation,
\begin{equation}
    \hat{\mathcal R}_{\ul{ABCD}}=\hat{R}_{\ul{ABCD}}+\hat K_{\ul{AC}}\hat K_{\ul{BD}}-\hat K_{\ul{BC}}\hat K_{\ul{AD}}\,,
\end{equation}
along with \eqref{eq:Riemann} and \eqref{eq:exCurv} to get
\begin{align}
    \hat{\mathcal R}_{\ul{\alpha\beta\gamma\delta}}&=\mathcal R_{\ul{\alpha\beta\gamma\delta}}-\frac{1}{2}F_{\ul{\alpha\beta}}F_{\ul{\gamma\delta}}-\frac{1}{2}F_{\ul{\alpha\vphantom{\beta}}[\ul\gamma|}F_{\ul\beta|\ul{\delta\vphantom{\beta}}]}\,,\nn\\
    \hat{\mathcal R}_{\ul{\alpha\beta\gamma}\ol z}&=-\frac{1}{2}\mathcal D_{\ul\gamma}F_{\ul{\alpha\beta}}\,,\nn\\
    \hat{\mathcal R}_{\ul{\alpha}\ol z\ul{\beta}\ol z}&=\frac{1}{4}F_{\ul{\alpha\gamma}}F_{\ul\beta}{}^{\ul\gamma}\,, \label{eq:indRiem}
\end{align}
where $\mathcal D$ again denotes the boundary covariant derivative. This underscores the fact that the induced Riemann tensor contains only intrinsic data about the boundary. In particular, note that \eqref{eq:indRiem} is exactly what we would expect if we forgot about the bulk and just reduced the boundary $\partial\mathcal M$ on a circle, 
$$\dd\sigma^2=h_{\mu\nu}\dd x^\mu\dd x^\nu+\qty(\dd z+A_{\ul\mu}\dd x^{\ul\mu})^2.$$
This is as expected from our discussion in Section \ref{sec:GHY}.

The induced Ricci tensor is thus
\begin{equation}
     \hat{\mathcal R}_{\ul{\alpha\beta}}=\mathcal R_{\ul{\alpha\beta}}-\frac{1}{2}F_{\ul{\alpha\gamma}}F_{\ul\beta}{}^{\ul\gamma}\,,\qquad\hat{\mathcal R}_{\ul{\alpha}\ol z}=\frac{1}{2}\mathcal D^{\ul\gamma}F_{\ul{\alpha\gamma}}\,,\qquad\hat{\mathcal R}_{\ol{zz}}=\frac{1}{4}F_{\ul{\alpha\beta}}F^{\ul{\alpha\beta}}\,,
\end{equation}
and the induced Ricci scalar is
\begin{equation}
    \hat{\mathcal R}=\mathcal R-\frac{1}{4}F_{\ul{\alpha\beta}}F^{\ul{\alpha\beta}}\,.
\end{equation}
Thus, the GHY term \eqref{eq:GBGHYinit} becomes
\begin{align}
    S_{\partial\mathrm{GB}}&=\frac{1}{\kappa_d}\int_{\partial\mathcal B}\dd[d-1]x\sqrt{-h}\Big[\mathcal Q_3+2K^{\ul{\alpha\beta}}F_{\ul{\alpha\gamma}}F_{\ul{\beta}}{}^{\ul\gamma}-\frac{1}{2}KF_{\ul{\alpha\beta}}F^{\ul{\alpha\beta}}-2\mathcal D^{\ul\gamma}F_{\ul{\alpha\gamma}}F_{\delta\ul{\alpha\vphantom{\beta}}}n^\delta\nn\\
    &\kern10em\!-K^{\ul{\alpha\beta}}F_{\gamma\ul{\alpha\vphantom{\beta}}}F_{\delta\ul\beta}n^\gamma n^\delta+KF_{\gamma\alpha}F_{\delta}{}^{\alpha}n^\gamma n^\delta\Big]\,.\label{eq:GBGHYred}
\end{align}
Combining \eqref{eq:GBGHYred} with the boundary term in \eqref{eq:GB}, and suitably integrating by parts, we get an effective lower-dimensional GHY term
\begin{align}
    \frac{1}{\kappa_d}\int_{\partial\mathcal B}\dd[d-1]x\sqrt{-h}\qty[\mathcal Q_3+\qty(K_{\ul{\alpha\beta}}-\frac{1}{2}Kh_{\ul{\alpha\beta}})F^{\ul{\alpha\gamma}}F^{\ul\beta}{}_{\ul\gamma}]\,.\label{eq:GHYGBred2}
\end{align}
In particular, $\mathcal Q_3$ is the GHY term corresponding to Gauss-Bonnet, so the remaining piece must be associated with the $RFF$ term. Indeed, we observe that this term may be written as
\begin{equation}
    -\frac{1}{2}R_{\alpha\beta\gamma\delta}F^{\alpha\beta}F^{\gamma\delta}+2R_{\alpha\beta}F^{\alpha\gamma}F^\beta{}_{\gamma}-\frac{1}{2}RF^2=-\frac{1}{2}R_{\alpha\beta}{}^{\gamma\delta}\qty(F^{\alpha\beta}F_{\gamma\delta}-4\delta^{[\alpha}_{[\gamma}F^{\beta]\epsilon}F_{\delta]\epsilon}+\delta^{[\alpha}_{[\gamma}\delta^{\beta]}_{\delta]}F^2)\,.
\end{equation}
We may then pull out the same combination of field strengths from the GHY term, \eqref{eq:GHYGBred2}. In particular, the tensors we have available to work with are $K_{\alpha\beta}$, $h_{\alpha\beta}$, and $n_\alpha$. Thus, we see that the GHY term may be rewritten as
\begin{equation}
    \frac{1}{\kappa_d}\int_{\partial\mathcal B}\dd[d-1]x\sqrt{-h}\qty[\mathcal Q_3+\qty(F^{\alpha\beta}F_{\gamma\delta}-4\delta^{[\alpha}_{[\gamma}F^{\beta]\epsilon}F_{\delta]\epsilon}+\delta^{[\alpha}_{[\gamma}\delta^{\beta]}_{\delta]}F^2)n_\alpha n_\delta K_{\beta\gamma}]\,.\label{eq:GBghy}
\end{equation}
In particular, this suggests that, in general, a term of the form $U^{\alpha\beta\gamma\delta}R_{\alpha\beta\gamma\delta}$, where $U^{\alpha\beta\gamma\delta}$ has the same symmetries as the Riemann tensor and contains no derivatives of the metric, will have an associated GHY term of $-4 U^{\alpha\beta\gamma\delta}n_\alpha n_\delta K_{\beta\gamma}$. Indeed, this is the case, as was originally shown in \cite{Barvinsky:1995dp}.

Notice that the particular combination appearing in \eqref{eq:GBghy} means that the normal components of $F$ precisely cancel. As a result, we do not have to introduce additional boundary conditions for $\delta(n^\mu F_{\mu\nu})$, and we continue to have a well-defined Dirichlet problem.\footnote{Since $\delta A_\mu$ vanishes everywhere on $\partial\mathcal B$, the tangential derivative must also vanish. Thus, having tangential components of the field strength does not adversely affect the consistency of the variational principle.} This is because we started with the Gauss-Bonnet action, for which the Dirichlet boundary conditions are sufficient, and this continues to hold after dimensional reduction. More generally, the GHY term $-4 U^{\alpha\beta\gamma\delta}n_\alpha n_\delta K_{\beta\gamma}$ will be sufficient to allow Dirichlet boundary conditions for the metric, but may create problems for other fields appearing in $U$, such as $A_{\mu}$. However, at least for the case of a gauge field, this could presumably be fixed by the addition of a Hawking-Ross term \cite{Hawking:1995ap}.

\subsection{Mixed Chern-Simons term}
We have seen that we can reduce an action for which we know the GHY term to obtain GHY terms for the reduced action. We can also go in reverse. That is, we can start with a lower-dimensional action and infer what must generate that GHY term upon reduction. We will demonstrate this for the case of the mixed Chern-Simons term in 5d:
\begin{equation}
    S_\mathrm{CS}^{(5)}=\frac{1}{2\kappa_5}\int_{\mathcal M}\,\Tr[\hat R\land \hat R]\land\hat A_{(1)}\,,
\end{equation}
where
\begin{equation}
    \Tr[\hat R\land \hat R]=\hat R^{AB}\land\hat R_{BA}=\frac{1}{4}\hat R_{MNAB}R_{PQ}{}^{BA}\dd \hat x^M\land\dd \hat x^N\land\dd \hat x^P\land\dd \hat x^Q\,,
\end{equation}
and $\hat A_{(1)}$ is a one-form field. Terms such as these notably appear in the action of $\mathcal N=2$, $D=5$ supergravity \cite{Myers:2009ij}. The appropriate GHY term is not known. However, we may perform a restricted circle reduction
\begin{equation}
    \dd s^2=g_{\mu\nu}\dd x^\mu\dd x^\nu+\dd z^2,\qquad \hat A=\phi\,\dd z\,.\label{eq:babyKK}
\end{equation}
This results in the 4d Chern-Simons modified gravity (see \emph{e.g.} \cite{Jackiw:2003pm})
\begin{equation}
    S_\mathrm{CS}^{(4)}=\frac{1}{2\kappa_4}\int_{\mathcal B}\Tr[\hat R\land \hat R]\,\phi\, ,
\end{equation}
which is just a scalar coupled to the Chern-Pontryagin term. This is known to require a GHY term \cite{Grumiller:2008ie}
\begin{equation}
    S_\mathrm{\partial CS}^{(4)}=\frac{2}{\kappa_4}\int_{\partial\mathcal B}\mathrm{CS}\qty(K)\,\phi\,,\qquad \mathrm{CS}(K)=\Tr[K\land\mathcal D K]=\frac{1}{2}K_\mu{}^\alpha\mathcal D_\nu K_{\rho\alpha}\dd x^\mu\land\dd x^\nu\land\dd x^\rho\,,\label{eq:4dCSGHY}
\end{equation}
which, notably, depends on the third fundamental form. This is precisely the boundary term required for the Hirzebruch signature index theorem \cite{Eguchi:1980jx}.

In particular, the reduction \eqref{eq:babyKK} is quite simple: The extrinsic curvature and boundary covariant derivatives reduce trivially as
\begin{equation}
    \hat K_{\alpha\beta}=K_{\alpha\beta}\,,\qquad \hat K_{\alpha\ol z}=K_{\ol{zz}}=0\,,\qquad \hat{\mathcal D}_\alpha \hat K_{\beta\gamma}=\mathcal D_\alpha K_{\beta\gamma}\,.
\end{equation}
This implies that the 4d GHY term \eqref{eq:4dCSGHY} should arise from a 5d GHY term
\begin{equation}
    S_\mathrm{\partial CS}^{(5)}=\frac{2}{\kappa_5}\int_{\partial\mathcal M}\mathrm{CS}(\hat K)\land\hat A_{(1)}\,.\label{eq:bCSa}
\end{equation}
It is important to emphasize that if this did not yield a well-posed variational principle in 5D, it would lead to an ill-posed one in 4D. Since this is not the case, we can be confident that \eqref{eq:bCSa} gives a well-posed variation.

More generally, one might consider the $D$-dimensional action
\begin{equation}
    S_\mathrm{CS}^{(D)}=\frac{1}{2\kappa_D}\int_{\mathcal M}\,\Tr[\hat R\land \hat R]\land\hat A_{(D-4)}\,,
\end{equation}
where $\hat A_{(D-4)}$ is a $(D-4)$-form. Using a torus reduction
\begin{equation}
    \dd s^2=g_{\mu\nu}\dd x^\mu\dd x^\nu+\delta_{ij}\dd z^i\dd z^j\,,\qquad A_{(D-4)}=\phi\,\dd z^1\land\cdots\land\dd z^{D-4}\,,
\end{equation}
it is clear that the corresponding GHY term should be
\begin{equation}
    S_\mathrm{\partial CS}^{(D)}=\frac{2}{\kappa_D}\int_{\partial\mathcal M}\mathrm{CS}(\hat K)\land\hat A_{(D-4)}\,.
\end{equation}
This result is not so surprising, and we could have guessed this. The interesting aspect is that the uplifting procedure is rigorous. So we get the GHY term directly, rather than a guess that we must check by varying the action, although we verify this in Appendix \ref{app:ARR} for completeness.

\subsection{$f(R)$ gravity}
Let us now consider the case of $f(R)$ gravity. The bulk action is given by~\cite{Sotiriou:2008rp}
\begin{equation}
    S_{f(R)}=\frac{1}{2\kappa_D}\int_{\mathcal M}\dd[D]x\sqrt{-\hat g}f(\hat R)\,,\label{eq:f(R)}
\end{equation}
where $f(\hat R)$ is a smooth function of $\hat R$. The corresponding GHY term is simply given by~\cite{Barth:1984jb,Madsen:1989rz,Nojiri:1999nd,Casadio:2001ff,Balcerzak:2008bg}
\begin{equation}
    S_{\partial f(R)}=\frac{1}{\kappa_D}\int_{\partial\mathcal M}\dd[D-1]x\sqrt{-\hat h}f'(\hat R)\hat K\,.
\end{equation}
Note that, in general, this does not lead to second-order equations of motion, except for the special case $f(\hat R)=\hat R-2\Lambda$. As a result, it is insufficient to specify Dirichlet boundary conditions alone,\footnote{However, Dirichlet boundary conditions are sufficient if one restricts to maximally symmetric backgrounds \cite{Madsen:1989rz}.}
and we must fix the additional degrees of freedom. In particular, we will have a good variational principle so long as $\delta(f'(\hat R))$ vanishes on the boundary. This is most naturally accomplished by
\begin{equation}
    \delta(f'(\hat R))\big\vert_{\partial\mathcal M}=f''(\hat R)\delta\hat R\big\vert_{\partial\mathcal M}=0\implies\delta\hat R\big\vert_{\partial\mathcal M}=0\,.\label{eq:Rboundary}
\end{equation}
As a result, $f(\hat R)$ theories with $f''(\hat R)\ne 0$ have an extra degree of freedom compared with two-derivative Einstein gravity. The condition \eqref{eq:Rboundary} may seem a bit unsettling at first glance; however, this approach suffices to correctly reproduce the ADM energy and the semiclassical Wald entropy~\cite{Dyer:2008hb}. Alternatively, it is well known that \eqref{eq:f(R)} is classically equivalent to an $\omega=0$ Brans-Dicke theory with action~\cite{Teyssandier:1983zz,Whitt:1984pd,Barrow:1988xh,Barrow:1988xi,Wands:1993uu,Chiba:2003ir}
\begin{equation}
    S_\mathrm{BD}=\frac{1}{2\kappa_D}\int_{\mathcal M}\dd[D]x\sqrt{-\hat g}\qty[f(\varphi)+f'(\varphi)(\hat R-\varphi)]\,,
\end{equation}
in which case the condition \eqref{eq:Rboundary} is just the Dirichlet boundary condition for the scalar, $\delta\varphi\vert_{\partial\mathcal M}=0$. 

It is now straightforward to reduce the bulk action \eqref{eq:f(R)}, again using our restricted circle reduction \eqref{eq:U1red}:
\begin{align}
    S_{f(R)}=\frac{1}{2\kappa_d}\int_{\mathcal B}\dd[d]x\sqrt{-g}\,f\qty(R-\frac{1}{4}F^2)\,.
\end{align}
We may expand this formally as a series
\begin{align}
    S_{f(R)}=\frac{1}{2\kappa_d}\int_{\mathcal B}\dd[d]x\sqrt{-g}\sum_{n=0}^\infty\frac{1}{n!}f^{(n)}(R)\qty(-\frac{1}{4}F^2)^n.
\end{align}
Reducing the GHY action, we get
\begin{equation}
    S_{\partial f(R)}=\frac{1}{\kappa_d}\int_{\partial\mathcal B}\dd[d-1]x\sqrt{-h}\,f'\qty(R-\frac{1}{4}F^2)K\,,
\end{equation}
which may likewise be expanded as a power series
\begin{equation}
    S_{\partial f(R)}=\frac{1}{\kappa_d}\int_{\partial\mathcal B}\dd[d-1]x\sqrt{-h}\sum_{n=0}^\infty\frac{1}{n!} f^{(n+1)}\qty(R)\qty(-\frac{1}{4}F^2)^nK\,.
\end{equation}
Hence, matching term by term,\footnote{We expect that the GHY term leading to a well-posed variation of the metric has the same power of field strengths as the bulk term, else the metric variation would not be properly controlled.} we see that combinations of the form
\begin{equation}
    \frac{1}{2\kappa_d}\int_{\mathcal B}\dd[d]x\sqrt{-g}\,f^{(n)}(R)\qty(F^2)^n+\frac{1}{\kappa_d}\int_{\partial\mathcal B}\dd[d-1]x\sqrt{-h}\,f^{(n+1)}(R)\qty(F^2)^nK\,,
\end{equation}
will have a well-posed variational problem so long as we fix appropriate boundary conditions. Now, recall that the choice of $f$ is arbitrary. We could, for example, choose $f$ to be the $n^\text{th}$ antiderivative of some other smooth function $f_n(R)$ so that
\begin{equation}
    \frac{1}{2\kappa_d}\int_{\mathcal B}\dd[d]x\sqrt{-g}\,f_n(R)\qty(F^2)^n+\frac{1}{\kappa_d}\int_{\partial\mathcal B}\dd[d-1]x\sqrt{-h}\,f_n'(R)\qty(F^2)^nK\,,
\end{equation}
has a well-posed variation. We could then sum over $n$ to obtain a new generalized action, where now the $f_n$ act as ($R$-dependent) Taylor series coefficients. That is, define a new function
\begin{equation}
    f(R,F^2):=\sum_n f_n(R) (F^2)^n\,.
\end{equation}
Since we have the freedom to choose the $f_n(R)$ to be any smooth functions, $f(R,F^2)$ is itself an arbitrary smooth function. After summing, we get the action
\begin{equation}
    S_{f(R,F^2)}=\frac{1}{2\kappa_d}\int_{\mathcal B}\dd[d]x\sqrt{-g}\,f(R,F^2)\,,
\end{equation}
with boundary term 
\begin{equation}
    S_{\partial f(R,F^2)}=\frac{1}{\kappa_d}\int_{\partial\mathcal B}\dd[d-1]x\sqrt{-h}\,\partial_R f(R,F^2)K\,.
\end{equation}
The above discussion automatically implies that this action will be well-posed, so long as we have suitable boundary conditions.

Before generalizing, we would expect to have
\begin{equation}
    \delta\qty(R-\frac{1}{4}F^2)\Bigg\vert_{\partial\mathcal B}=0\,,\label{eq:RF2case}
\end{equation}
which is the dimensional reduction of \eqref{eq:Rboundary}. However, in Taylor expanding and resumming, we have generalized the functional dependence on $R$ and $F^2$, so we should expect a generalization of \eqref{eq:RF2case}. Indeed, we must impose the boundary condition
\begin{equation}
    \delta \qty(\partial_R f(R,F^2))\big\vert_{\partial\mathcal B}=0\,,\label{eq:fRredBCs}
\end{equation}
which is directly analogous to \eqref{eq:Rboundary}. This can be obtained by dimensionally reducing, Taylor expanding, and resumming \eqref{eq:Rboundary}, since
\begin{align}
    \delta f'(\hat R)\big\vert_{\partial\mathcal M}=0&\implies\delta f'\qty(R-\tfrac{1}{4}F^2)\big\vert_{\partial\mathcal B}=0\implies \partial_R\delta\qty(f^{(n)}(R)(F^2)^n)\Big\vert_{\partial\mathcal B}=0\nn\\
    &\implies 0=\partial_R\delta\qty(\sum_n f_n(R)(F^2)^n)\Bigg\vert_{\partial\mathcal B}=\delta\qty(\partial_R f(R,F^2))\big\vert_{\partial\mathcal B}\,.
\end{align}
In Appendix \ref{app:frf2}, we rigorously check that the boundary condition \eqref{eq:fRredBCs} is necessary and sufficient.

Notice that if $\partial_R\partial_{F^2}f(R,F^2)=0$, then the condition \eqref{eq:fRredBCs} reduces to that of $f(R)$ gravity,
\begin{equation}
    \partial_R\partial_{F^2}f(R,F^2)=0\implies\delta \qty(\partial_R f(R,F^2))=\partial_R^2f(R,F^2)\,\delta R\implies \delta R\big\vert_{\partial\mathcal B}=0\,.
\end{equation}
This makes sense since $\partial_R\partial_{F^2}f(R,F^2)=0$ implies that there are no cross terms between $R$ and $F^2$, hence the Lagrangian decouples into an $f(R)$ piece and a matter Lagrangian. In the case that $f(R,F^2)=f(R-\tfrac{1}{4}F^2)$, \eqref{eq:fRredBCs} indeed reduces to \eqref{eq:RF2case}.

 If the only matter content is a Maxwell field, then $F^2$ is proportional to the trace of the stress-energy tensor, $T$. This suggests a further generalization to $f(R,T)$ gravity (see \emph{e.g.}~\cite{Harko:2011kv}):
\begin{equation}
    S_{f(R,T)}=\frac{1}{2\kappa_d}\int_{\mathcal B}\dd[d]x\sqrt{-g}\,\qty[f(R,T)+\mathcal{L}{}_{\mathrm{matter}}]\,,
\end{equation}
which should have a GHY term
\begin{equation}
    S_{\partial f(R,T)}=\frac{1}{\kappa_d}\int_{\partial\mathcal B}\dd[d-1]x\sqrt{-h}\,\partial_R f(R,T)K\,,
\end{equation}
subject to the boundary condition
\begin{equation}
    \delta\qty(\partial_Rf(R,T))\big\vert_{\partial\mathcal B}=0\,.
\end{equation}
This is straightforward to check, as the proof is identical to that for the case of the $f(R,F^2)$ action. 

It should also be noted that this applies to the $f(R,\mathcal L_\mathrm{matter})$ theory~\cite{Harko:2010mv} as well, or even more generally the $f(R,T,\mathcal L_\mathrm{matter})$ theory~\cite{Haghani:2021fpx}. Since we never used any details of the stress-energy tensor, it should be clear that
\begin{equation}
    \frac{1}{2\kappa_d}\int_{\mathcal B}\dd[d]{x}\sqrt{-g}f(R,T,\mathcal L_\mathrm{matter})+\frac{1}{\kappa_d}\int_{\partial\mathcal B}\dd[d-1]x\sqrt{-h}\,\partial_R f(R,T,\mathcal L_\mathrm{matter})K\,,
\end{equation}
also has a well-defined variational principle, subject to imposing the boundary conditions
\begin{equation}
    \delta\qty(\partial_Rf(R,T,\mathcal L_\mathrm{matter}))\big\vert_{\partial\mathcal B}=0\,.
\end{equation}
Again, the proof is omitted as it is nearly identical to that for $f(R,F^2)$ gravity.
    
\section{Discussion}\label{sec:disc}
We have investigated what happens to GHY terms upon dimensional reduction and found that this method is useful for producing GHY terms without having to work out any variations. In particular, we always get precisely the GHY terms of the reduced theory. This is in disagreement with the result of \cite{Kim:2024mhb}, which found that the circle reduction from 5d to 4d produces an additional abelian Chern-Simons term for the KK gauge field; however, it seems to be the case that the result of~\cite{Kim:2024mhb} differs because the authors neglected to consider the boundary term \eqref{eq:bdyTerm} when integrating the bulk action by parts. As a result, the scalar derivative remained uncancelled, which, after the use of the equations of motion, resulted in the seeming appearance of an abelian Chern-Simons term.

It is worth noting that nothing prevented the four-derivative analysis from being applied to a full group manifold reduction, such as in \eqref{eq:gpMfldRed}, keeping all the scalars. However, it is considerably more technically involved and it does not seem that much insight would be gained from including the scalars. In particular, the GHY terms do not affect the equations of motion,\footnote{Although not relevant to the current work, it should be noted that boundary terms can affect the equations of motion in ADM reductions, see \emph{e.g.} \cite{Park:2013vpa,Park:2013bma}.} so our current investigation is orthogonal to the usual worries about the consistency of truncations.

In some sense, this work is a proof of principle. It would be interesting to see if more diverse and exciting GHY terms can be obtained using this prescription. In particular, the GHY surface terms for the Lovelock lagrangians \cite{Lovelock:1970zsf,Lovelock:1971yv}
\begin{equation}
    \mathcal{X}_{2n}=\frac{1}{2^n}\delta^{\mu_1\cdots\mu_{2n}}_{\nu_1\cdots\nu_{2n}}R^{\nu_1\nu_2}_{\mu_1\mu_2}\cdots R^{\nu_{2n-1}\nu_{2n}}_{\mu_{2n-1}\mu_{2n}}\,,
\end{equation}
are known to be given by \cite{Myers:1987yn,Teitelboim:1987zz},
\begin{equation}
    \mathcal{Q}_{2n-1}=2n\int_0^1\dd t\,\delta^{\ul{\alpha}_1\cdots\ul\alpha_{2n-1}}_{\ul{\beta}_1\cdots\ul\beta_{2n-1}}K^{\ul\beta_1}_{\ul\alpha_1}\qty[\frac{1}{2}\mathcal R^{\ul\beta_2\ul\beta_3}_{\ul\alpha_2\ul\alpha_3}-t^2K^{\ul\beta_2}_{\ul\alpha_2}K^{\ul\beta_3}_{\ul\alpha_3}]\cdots\qty[\frac{1}{2}\mathcal R^{\ul\beta_{2n-2}\ul\beta_{2n-1}}_{\ul\alpha_{2n-2}\ul\alpha_{2n-1}}-t^2K^{\ul\beta_{2n-2}}_{\ul\alpha_{2n-2}}K^{\ul\beta_{2n-1}}_{\ul\alpha_{2n-1}}]\,.
\end{equation}
In principle, our procedure should then be able to extract the GHY terms for terms of the form $R^2FF$. We expect they would show up in a combination akin to \eqref{eq:GBghy}, where the normal components of $F$ cancel, but with sufficient cleverness, one should be able to extract the general form, as in \eqref{eq:GBghy}.

A second application of the present work is an $f(\mathcal{X}_4)$ term. This action and its corresponding GHY term are given by~\cite{Bueno:2016dol}
    \begin{equation}
        \frac{1}{2\kappa_D}\qty[\int_{\mathcal M}\dd[D]x\sqrt{-g}\,f(\mathcal{X}_4)+\int_{\partial\mathcal M}\dd[D-1]x\sqrt{-h}\,f'(\mathcal{X}_4)\mathcal Q_3]\,,
    \end{equation}
which must then be supplemented by the extra boundary condition
\begin{equation}
    \delta\mathcal X_4\big\vert_{\partial\mathcal M}=0\,,
\end{equation}
analogous to Eq. \eqref{eq:Rboundary} for $f(R)$ gravity. In principle, we have already reduced $\mathcal X_4$ and $\mathcal Q_3$, so there is no obstruction to writing down the reduced action. However, it is more complicated to deal with the leftover derivatives. It would be interesting to see if any new insights can be gained.

Finally, one might ask if something similar can be done for counterterms. Counterterms are required to have a finite action and are important for holographic renormalization \cite{Bianchi:2001kw}, and holographically correspond to the usual CFT counterterms. However, they are also more subtle than the GHY terms. For example, two-derivative gauged supergravity in $D$ dimensions requires a counterterm~\cite{Batrachenko:2004fd}
\begin{equation}
    \frac{1}{2\kappa_D}\int_{\partial\mathcal M}\dd[D-1]x\sqrt{-h}\qty[W(\phi)+\frac{L}{(D-3)}\mathcal R+\frac{L^3}{(D-3)^2(D-5)}\qty(\mathcal R_{\ul{\alpha\beta}}^2-\frac{D-1}{4(D-2)}\mathcal R^2)]\,,
\end{equation}
where $L$ is the inverse AdS radius, and the terms above are sufficient for $D\le 6$. In particular, the coefficients are all dependent on the dimension and the leading term $W(\phi)$ depends on the matter content of the theory. We leave this problem to future work.

\section*{Acknowledgements}
I would like to thank Jim Liu for comments on the draft. This work was supported in part by the U.S. Department of Energy under grant DE-SC0007859, in part by a Leinweber Summer Research Award, and in part by the National Key Research and Development Program No. 2022YFE0134300.

\appendix
\section{Submanifolds}\label{app:math}
Here, we summarize some useful mathematics regarding codimension-one submanifolds, mostly following \cite{Speranza:2019hkr}. Viewing the spacetime as the normal bundle over the boundary, we write the metric as
\begin{equation}
    g_{\mu\nu}=h_{\mu\nu}+n_\mu n_\nu\,,
\end{equation}
where, after being appropriately pulled back, $n_\mu n_\nu$ is naturally identified as the metric of the normal space, and $h_{\mu\nu}$ is the induced metric on the boundary. Indices are raised and lowered using the bulk metric, $g_{\mu\nu}$. Thus, $n_\mu$ naturally forms an einbein for the normal space. Written in terms of a frame $e^\alpha$, we have
\begin{equation}
    \eta_{\alpha\beta}=h_{\alpha\beta}+n_\alpha n_\beta\,.
\end{equation}

The extrinsic curvature is given by
\begin{equation}
    K_{\alpha\beta}=h_\alpha{}^\gamma h_\beta{}^\delta \nabla_\gamma n_\delta\,,
\end{equation}
and so we may write
\begin{equation}
    n^\alpha K_{\beta\gamma}=n^\alpha h_\beta{}^\delta h_\gamma{}^\epsilon \nabla_\delta n_\epsilon=h_\beta{}^\delta h_\gamma{}^\epsilon \nabla_\delta \qty(n^\alpha n_\epsilon)=-h_\beta{}^\delta h_\gamma{}^\epsilon \nabla_\delta h_\epsilon{}^\alpha\,.\label{eq:Kidentity}
\end{equation}
Given a tangential vector $V^\alpha$ such that $n_\alpha V^\alpha=0$, the intrinsic derivative $\mathcal D$ is defined as the projection of the covariant derivative onto the boundary 
\begin{equation}
    \mathcal D_{\alpha}V^{\beta}\equiv h^{\gamma}{}_{\alpha}h^{\beta}{}_{\delta}\nabla_\gamma V^\delta=h^{\gamma}{}_{\alpha}\nabla_\gamma V^{\beta}+n^\beta K_{\alpha\delta}V^\delta,
\end{equation}
where the second equality follows from \eqref{eq:Kidentity}. This extends linearly to tensors with multiple tangential indices and is the unique covariant derivative compatible with the induced metric on the boundary, $\mathcal D_\alpha h_{\beta\gamma}=0$. The associated connection coefficients $\gamma^\mu_{\nu\rho}$ are precisely the Christoffel symbols for the induced metric $h$:
\begin{equation}
    \gamma^\mu_{\nu\rho}=\frac{1}{2}h^{\mu\sigma}\qty(\partial_\nu h_{\rho\sigma}+\partial_\rho h_{\nu\sigma}-\partial_\sigma h_{\nu\rho})\,.
\end{equation}
The associated curvature is the intrinsic Riemann curvature, in the sense that
\begin{equation}
    [\mathcal D_\gamma,\mathcal D_\delta]V^\alpha=\mathcal R^\alpha{}_{\beta\gamma\delta}V^\beta.
\end{equation}

The intrinsic derivative can be extended to act on vectors with normal indices $W^\alpha$, in the sense that $h^\alpha{}_\beta W^\beta=0$, by
\begin{equation}
    \mathcal D_{\alpha}W^{\beta}=h^{\gamma}{}_{\alpha}n^{\beta}n_{\delta}\nabla_\gamma W^\delta=h^{\gamma}{}_{\alpha}\nabla_\gamma W^{\beta}-n_\delta K_{\alpha}{}^{\beta}W^\delta,
\end{equation}
where the second equality again follows from \eqref{eq:Kidentity}. This makes $\mathcal D$ compatible with the normal metric $n_\alpha n_\beta$. The associated curvature is the outer curvature, which vanishes for a codimension-one submanifold:
\begin{equation}
    [\mathcal D_\alpha,\mathcal D_\beta]W^\gamma=0\,.
\end{equation}
This is just the statement that the normal space is one-dimensional and hence cannot have any non-trivial curvature.

For tensors with both normal and tangential indices, we act with $h_\alpha{}^\beta\nabla_\beta$ and tangentially project indices that were originally tangential and normally project indices that were originally normal. $\mathcal D$ then acts on indices that are not specifically tangential or normal by first projecting onto the tangential and normal components and then acting according to the preceding definitions. In particular,
\begin{equation}
    \mathcal D_\alpha g_{\beta\gamma}=\mathcal D_\alpha( h_{\beta\gamma}+n_\beta n_\gamma)=0\,.
\end{equation}

\section{Proofs of variations}
Here, we present several proofs that our proposed GHY terms do indeed lead to a well-posed variational problem.

\subsection{$\Tr[R\land R]\land A$}\label{app:ARR}
Consider the action
\begin{equation}
    S=\int_{\mathcal M}\Tr[R\land R]\land A_{(D-4)}+4\int_{\partial\mathcal M}\mathrm{CS}(K)\land A_{(D-4)}\,,
\end{equation}
subject to Dirichlet boundary conditions,
\begin{equation}
    \delta g_{\mu\nu}\big\vert_{\partial\mathcal M}=0\,.
\end{equation}
For the purposes of verifying the variational principle is well-posed, it is easiest to work with this action expressed in local coordinates:\footnote{Note we have rescaled by an overall factor of $(D-4)!$.}
\begin{align}
    S&=\frac{1}{4}\int_{\mathcal M}\dd[D]x\sqrt{-g}\,\epsilon^{\mu_1\cdots\mu_D}R_{\mu_1\mu_2\rho\sigma}R_{\mu_3\mu_4}{}^{\sigma\rho}A_{\mu_5\cdots\mu_D}\nn\\
    &\quad +2\int_{\partial\mathcal M}\dd[D-1]x\sqrt{-h}\,\epsilon^{\mu_1\cdots\mu_D}n_{\mu_1}K_{\ul\mu_2}{}^{\ul\nu}\mathcal D_{\ul\mu_3}K_{\ul{\mu_4\nu}}A_{\mu_5\cdots\mu_D}\,.
\end{align}
Varying this action with respect to the metric yields~\cite{Grumiller:2008ie}
\begin{align}
    \delta S&=\frac{1}{2}\int_{\mathcal M}\dd[D]x\sqrt{-g}\,\epsilon^{\mu_1\cdots\mu_D}R_{\rho\sigma\mu_1\mu_2}\delta R^{\sigma\rho}{}_{\mu_3\mu_4}A_{\mu_5\cdots\mu_D}\nn\\
    &\quad +4\int_{\partial\mathcal M}\dd[D-1]x\sqrt{-h}\,\epsilon^{\mu_1\cdots\mu_D}n_{\mu_1}\delta K_{\ul\mu_2}{}^{\ul\nu}\mathcal D_{\ul\mu_3}K_{\ul{\mu_4\nu}}A_{\mu_5\cdots\mu_D}\,.
\end{align}
up to irrelevant terms. We have
\begin{equation}
    \delta R^\mu{}_{\nu\rho\sigma}=\nabla_\rho\delta\Gamma^\nu_{\mu\sigma}-\nabla_\sigma\delta\Gamma^\nu_{\mu\rho}\,,
\end{equation}
and
\begin{equation}
    \delta\Gamma^\nu_{\mu\rho}=\frac{1}{2}g^{\nu\sigma}\qty(\nabla_\mu\delta g_{\sigma\rho}+\nabla_\rho\delta g_{\mu\sigma}-\nabla_\sigma\delta g_{\mu\rho})\,.
\end{equation}
Hence, integrating by parts yields
\begin{align}
    \delta S&=-\int_{\mathcal M}\dd[D]x\sqrt{-g}\,\epsilon^{\mu_1\cdots\mu_D}\nabla_{\mu_3}\qty(R_{\rho\sigma\mu_1\mu_2}A_{\mu_5\cdots\mu_D})\delta \Gamma^{\rho}_{\sigma\mu_4}\nn\\
    &\quad +\int_{\partial\mathcal M}\dd[D-1]x\sqrt{-h}\,\epsilon^{\mu_1\cdots\mu_D}n_{\mu_1}\qty(-2 R^{\rho\sigma}{}_{\mu_2\mu_3}n_\sigma n^\lambda\nabla_\lambda\delta g_{\rho\mu_4}+4\delta K_{\ul\mu_2}{}^{\ul\nu}\mathcal D_{\ul\mu_3}K_{\ul{\mu_4\nu}})A_{\mu_5\cdots\mu_D}\,,
\end{align}
where we have dropped the tangential derivatives of the metric variation since these necessarily vanish on the boundary. The first term needs to be further integrated by parts to obtain the equations of motion, but it will not affect the boundary term so we leave it alone. Finally, making use of the Codazzi equation
\begin{equation}
   n^\mu R_{\mu\ul{\nu\rho\sigma}}=\mathcal D_{\ul{\sigma\vphantom{\rho}}}K_{\ul{\nu\rho}}-\mathcal D_{\ul\rho}K_{\ul{\nu\sigma\vphantom{\rho}}}\,,
\end{equation}
along with 
\begin{equation}
    \delta K_{\ul{\mu\nu}}=\frac{1}{2}h_\mu{}^\rho h_\nu{}^\sigma n^\lambda\nabla_\lambda\delta g_{\rho\sigma}\,.
\end{equation}
we see that the boundary term vanishes.

\subsection{$f(R,F^2)$}\label{app:frf2}
Consider the action
\begin{equation}
    S=\int_{\mathcal M}\dd[D]x\sqrt{-g}\,f(R,F^2)+2\int_{\partial\mathcal M}\dd[D-1]x\sqrt{-h}\,\partial_Rf(R,F^2)K\,,
\end{equation}
subject to the boundary conditions
\begin{equation}
    \delta g_{\mu\nu}\big\vert_{\partial\mathcal B}=0\,,\qquad \delta A_{\mu}\big\vert_{\partial\mathcal B}=0\,,\qquad \delta\qty(\partial_Rf(R,F^2))\big\vert_{\partial\mathcal B}=0\,.\label{eq:frf2BC}
\end{equation}
The Dirichlet condition on the metric automatically implies that $\delta h_{\ul{\mu\nu}}=0$. Varying this action results in
\begin{align}
    \delta S=&\,\int_{\mathcal M}\dd[D]x\sqrt{-g}\Big[-\frac{1}{2}g^{\mu\nu}\delta g_{\mu\nu}f(R,F^2)+\partial_R f(R,F^2)\qty(R^{\mu\nu}\delta g_{\mu\nu}+g^{\mu\nu}\delta R_{\mu\nu})\nn\\
    &\kern6em\,\,\,+2\partial_{F^2}f(R,F^2)F^\mu{}_\rho F^{\nu\rho}\delta g_{\mu\nu}+4\partial_{F^2}f(R,F^2)F^{\mu\nu}\partial_\mu\delta A_\nu\Big]\nn\\
    &+2\int_{\partial\mathcal M}\dd[D-1]x\sqrt{-h}\qty[\delta\qty(\partial_Rf(R,F^2))+\partial_R f(R,F^2)\delta K]\,.\label{eq:frvar}
\end{align}
It is straightforward to compute
\begin{equation}
    \delta R_{\mu\nu}=\nabla_\rho\delta\Gamma^\rho_{\mu\nu}-\nabla_\nu\delta\Gamma^\rho_{\mu\rho}\,,
\end{equation}
and so, using the definition of the Christoffel symbols,
\begin{equation}
    g^{\mu\nu}\delta R_{\mu\nu}=g^{\mu\nu}\nabla^\rho\qty(\nabla_\nu\delta g_{\mu\rho}-\nabla_\rho\delta g_{\mu\nu})\,.
\end{equation}
We may also compute
\begin{equation}
    \delta K=\frac{1}{2}n^\gamma h^{\ul{\mu\nu}}\nabla_\gamma \delta g_{\mu\nu}\,.
\end{equation}
Now we plug this back into \eqref{eq:frvar}, integrate by parts, and use the generalized Stokes theorem to get
\begin{align}
    \delta S=&\,\int_{\mathcal M}\dd[D]x\sqrt{-g}\Bigg[\bigg(\partial_Rf(R,F^2)R^{\mu\nu}-\nabla^\mu\nabla^\nu\partial_R f(R,F^2)+\Box\partial_Rf(R,F^2)g^{\mu\nu}-\frac{1}{2}g^{\mu\nu}f(R,F^2)\nn\\
    &\kern7em+2\partial_{F^2}f(R,F^2)F^\mu{}_\rho F^{\nu\rho}\bigg)\delta g_{\mu\nu}-4\nabla_\mu\qty(\partial_{F^2}f(R,F^2)F^{\mu\nu})\delta A_\nu\Bigg]\nn\\
    &+2\int_{\partial\mathcal M}\dd[D-1]x\sqrt{-h}\Big[\delta\qty(\partial_Rf(R,F^2))+4\partial_{F^2}f(R,F^2)F^{\mu\nu}n_\mu\delta A_\nu\Big]\,,
\end{align}
where we have already discarded terms that depend on $\delta g_{\mu\nu}\vert_{\partial\mathcal M}$ and its tangential derivatives. We see that the first two lines are our equations of motion, and the third line vanishes per our boundary conditions, \eqref{eq:frf2BC}. Hence, the variational principle is well-posed.

\bibliographystyle{JHEP}
\bibliography{cite}

\end{document}